# Semiconductor laser mode locking stabilization with optical feedback from a silicon PIC

Johannes Hauck, Andrea Zazzi, Alexandre Garreau, François Lelarge,
Alvaro Moscoso-Mártir, Florian Merget, Jeremy Witzens, *Sr. Member, IEEE*

*Abstract*—Semiconductor mode-locked lasers can be used in a variety of applications ranging from multi-carrier sources for WDM communication systems to time base references for metrology. Their packaging in compact chip- or module-level systems remains however burdened by their strong sensitivity to back-reflections, quickly destroying the coherence of the mode-locking. Here, we investigate the stabilization of mode-locked lasers directly edge coupled to a silicon photonic integrated circuit, with the objective of moving isolators downstream to the output of the photonic circuit. A 2.77 kHz 3 dB RF linewidth, substantially improved compared to the 15.01 kHz of the free running laser, is obtained in the best case. Even in presence of detrimental reflections from the photonic circuit, substantial linewidth reductions from 20 kHz to 8.82 kHz, from 572 kHz to 14.8 kHz, and from 1.5 MHz to 40 kHz are realized.

## I. INTRODUCTION

Benchtop (Ti:sapphire) mode-locked lasers (MLLs) are able to produce pulse trains with extremely low phase noise [1], beating even the best radio-frequency (RF) oscillators including sapphire loaded microwave cavity oscillators. This has led, e.g., to the proposal of using them as reduced jitter clocks in optically enabled analog-to-digital converters (ADCs) [2]. While semiconductor MLLs do not currently feature such low phase noise pulse trains, relatively good operation has been achieved with help of long-delay external feedback by embedding a semiconductor optical amplifier (SOA) in a 6 meter long recirculating fiber loop [3]. Weak, fully passive optical feedback applied to a quantum (Q-)dash MLL has also proven conducive for improving both its optical and RF linewidths [4], i.e., reducing the phase noise of the underlying optical carrier as well as of the pulse train that would be obtained after applying dispersion compensation to the emitted comb [5]. While the improvement of the RF linewidth shown in [4], from 151 kHz down to 3 kHz, is not nearly sufficient for the aforementioned low jitter ADC applications, the demonstrated reduction of the optical linewidth from the 2-10 MHz range, typical for single section Fabry-Perot type MLLs [5], down to 100 kHz enables utilization of individual comb lines as optical carriers for coherent communications [6],[7]. Semiconductor MLLs have further been utilized to generate optical carriers for direct detection wavelength division multiplexed (WDM) transceivers. Here, mode-locking is merely used as a means to obtain sufficiently low relative intensity noise (RIN) via mode partition noise reduction [8]. Moreover, optical combs can be used for a variety of sensing applications such as dual comb spectroscopy [9] or optical ranging [10] in which a narrow RF linewidth is also advantageous.

Integration of MLLs into such systems by means of flip-chip bonding [11] is unfortunately difficult, as very weak levels of uncontrolled feedback are already sufficient to destroy the mode-locking. While stabilizing feedback has already been implemented with fiber based systems [12],[13], the goal here is to apply chip scale feedback based on silicon photonics (SiP) technology that can be co-integrated with the rest of the integrated transceiver system [14].

Notably, heterogeneous integration constitutes an alternative approach for integrating MLLs into a SiP platform, with the additional advantage of reducing parasitic reflections between the active waveguide and the remaining photonic integrated circuit (PIC) [15]. Together with the availability of reduced loss silicon delay lines, this has enabled low RF and optical linewidths [16], albeit the large cavity in [16] also resulted in a relatively small free spectral range (FSR) of 1 GHz – an important aspect to take into consideration when benchmarking devices, as achieved RF linewidths typically grow with the FSR. The advantage of integrating III-V materials with lower loss SiP waveguides can be intuitively understood from the RF linewidth and pulse jitter generally scaling together with the Schawlow-Townes optical linewidth limit of the laser [17]. While a high FSR can also be maintained with large laser cavities by incorporating an intra-cavity filter, as shown in a heterogeneously integrated platform in [18], both the FSR and the center wavelength of the filter need to be tightly controlled, placing constraints on passive and/or active waveguide group index and dispersion.

Our objective is to compensate detrimental effects from parasitic reflections occurring inside the transceiver chip, to improve MLL mode-locking characteristics, and to move the isolator to the output of the integrated transmitter chip when such heterogeneous integration is not available. Moreover, weak external feedback with the topology proposed in this paper maintains the high FSR required when high power per comb line is targeted [8] while circumventing the tight requirements on waveguide group index associated to other techniques such as intra-cavity filters or broadband external feedback, that arise when the length of delay lines is scaled up.

This work was supported in part by the European Commission under Grant 619591 and by the Deutsche Forschungsgemeinschaft (DFG) for project PACE (650602). A. Zazzi, A. Moscoso-Mártir, F. Merget and J. Witzens are with the Institute of Integrated Photonics (IPH) of RWTH Aachen University, Aachen,

52074 Germany (e-mail: jwitzens@iph.rwth-aachen.de). J. Hauck was with IPH and is now with DURAG GmbH, 22453 Hamburg, Germany. A. Garreau is with III-V Lab, 91767 Palaiseau, France and F. Lelarge was with III-V Lab and is now with Almae Technologies, 91460 Marcoussis, France.

## II. DESCRIPTION OF PIC AND LASER

Broadband reflection of the entire comb spectrum has proven to be conducive in free-space [4] and fiber [12],[13] based setups. Such feedback schemes have also been monolithically integrated, either directly in III-V laser chips [19],[20], or by means of heterogeneous integration [21],[22]. A big advantage of combining III-V gain materials with SiP resides, here too, in the ability of implementing relatively low loss and compact (spooled) delay lines, with lengths up to 4 cm combined with 3.934 mm long cavities in [21] and [22].

We anticipate the strong dispersion associated with integrated waveguides, as well as fabrication induced variability of the group index, to potentially become a problem as the length of the delay lines is increased. Indeed, ideally the phase of each back-reflected comb line should be dialed in to yield stabilizing feedback with a constructive interference condition. Difficulties associated to waveguide dispersion in case of broadband feedback, even in the case of moderately long delay lines (~8 mm), are discussed in details in [22]. Rather than implementing broadband group index and dispersion control, we opt for a scheme in which only two comb lines are fed back to the laser with narrowband reflectors, each adjustable with a tunable phase and intensity. While reflection of a single line only carries information on the phase of the underlying optical carrier, feeding back two lines also feeds back the RF phase (the phase of the pulse train) yielded by the beat note of the two lines. Selection of two comb lines from a master MLL and injection into a slave MLL has for example been previously shown to allow complete synchronization of the two comb sources [23]. It is thus expected that dual line self-injection also contributes to RF linewidth reduction.

We build on a PIC architecture previously used for the stabilization of distributed feedback lasers (DFBs) [24], adding a second on-chip optical path to provide feedback for a second comb line (schematic in Fig. 1), wherein it should be noted that component characteristics were reoptimized for the specific application targeted here, within the limits of the utilized technology platform. The resonant frequencies of both rings as well as the phase delays interposed in each optical feedback path were intended to be independently tunable, so as to ensure phase and amplitude of the PIC reflection coefficient can be independently set for both at a fixed wavelength. Laser frequency pulling has been previously investigated for this scheme in the simpler case of single frequency feedback applied to a DFB [24] and is expected to also play a role here. This is discussed in further details in the context of an MLL in the analysis of the experimental results.

From a control perspective, four phases have to be dialed in: Two to tune the resonant wavelengths of the rings, and two others to set the overall phase of the two reflection paths. From an actuation perspective, this is more straightforward than e.g. incorporating an intra-cavity filter. Here, only the phase at a single MLL line matters for each optical path (which can be tuned with a simple phase shifter). On the other hand, in the case of an intra-cavity comb line selection filter, the filter's FSR has to be a multiple of the MLL's FSR. Thus, the waveguide's group index and dispersion are also critical. Dynamically tuning the group index of a waveguide independently of the effective index at a fixed wavelength, so as to adjust a filter's FSR while at the same time maintaining its alignment to a given resonance, boils down to wideband control of the dispersion curve and is much more difficult [25].

Maintaining the alignment of the ring resonances with the targeted comb lines is also relatively straightforward, as the power transmitted to the drop port can be monitored. On the other hand, dialing in and maintaining the right feedback phases in the scheme proposed here poses control challenges from the perspective of error signal acquisition, as the only obvious feedback resides in the RF linewidth of the MLL, that cannot be recorded without complex electronics. While there might be correlations between emitted power levels and the phase of the feedback [24], particularly at the wavelengths at which feedback is being applied, this was not investigated here in the context of an MLL and it remains unclear whether such a simplified feedback mechanism could be used in a practical setting. These difficulties are shared with other forms of external feedback, such as broadband feedback as implemented in [22]. While here too the strength of the feedback could be straightforwardly monitored, the phase of the feedback, influencing both pulse shape and RF linewidth, was not directly monitored within the chip. In comparison, monitoring the spectral alignment of an intracavity filter as used in [18] is more straightforward, as a direct feedback signal can be found in the form of the power transmitted through the filter, or even in the laser output power.

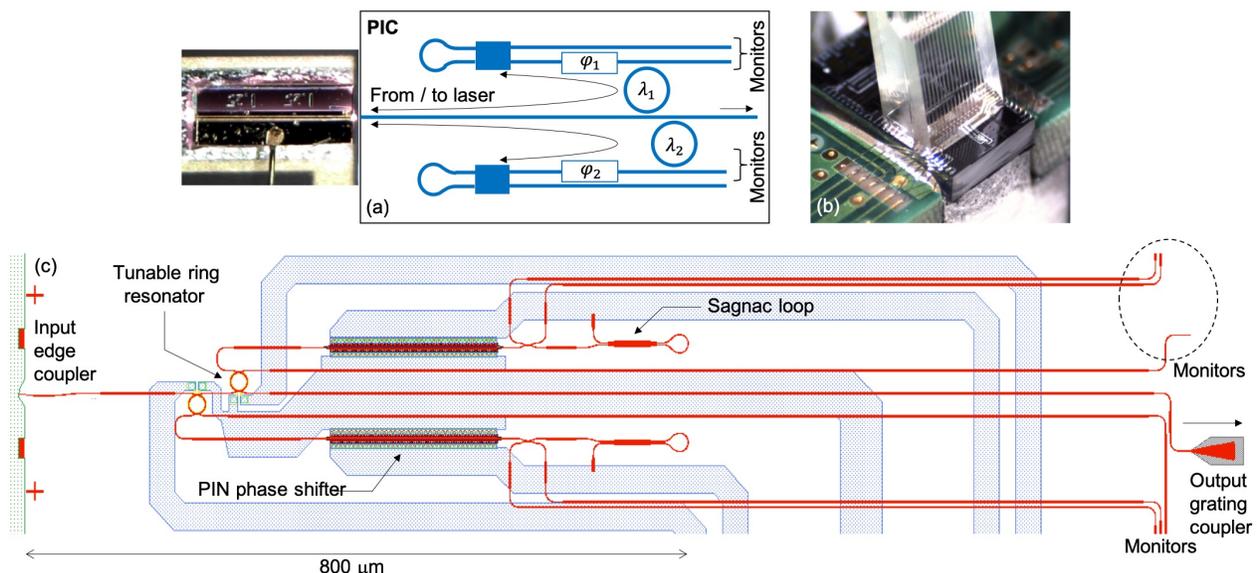

Fig. 1. (a) Schematic of PIC with two feedback paths tuned to wavelengths $\lambda_1$ and $\lambda_2$. The phase delays $\varphi_1$ and $\varphi_2$ are also tunable. (b) Photograph of PIC with mounted fiber array. The laser (not shown) is later positioned in front. (c) Detailed PIC layout with characteristic dimensions. Waveguides are represented in red and metal interconnects in dotted blue. The green area on the left shows the deep etch used to define the chip facet.

A theoretical investigation of a semiconductor MLL under self-injection feedback from two external cavities has been reported in [26], numerically showing the reduction of the pulse train's timing jitter, related to the RF linewidth [27], under constructive feedback conditions. The timing jitter is modeled to scale as the inverse of the time delay applied to the feedback [26], within the limits of the laser remaining a stable system, similarly to the optical linewidth reduction of a single mode laser under optical self-injection feedback [28]. However, a major difference between the system modeled in [26] and the experimental work pursued here resides in [26] assuming the FSR of the external cavities to coincide with the FSR of the laser (more precisely, the FSR of the laser being a multiple of the FSR of the external cavities). Consequently, the external cavity is assumed to constructively feed back the entire comb to the laser. This model is closer to broadband feedback and implies similar constraints in terms of group index and dispersion control.

In our case, the external cavities simply serve to each select one comb line, and their FSR is irrelevant so long as it is large enough for adjacent resonances not to substantially overlay with the MLL spectrum. As such, it is analogous to the work described in [29], in which a single-wavelength semiconductor laser is controlled by filtered feedback and subjected to multiple reflection paths. Beyond selecting a single line for the application of optical feedback, the external resonators can be simply modeled as applying an additional time (group) delay, as well as a tunable attenuation and phase delay, to the feedback path [24]. In particular, as described in the following, their linewidth is too large for direct shaping of the selected comb lines' optical spectrum. Their primary function inside the PIC is to allow for variable optical attenuation of the fed back comb lines.

### A. Detailed Description of the PIC

Figure 1(c) shows a detailed schematic of the PIC, which was implemented in the standard SiP platform of Singapore's Institute of Microelectronics (IME A*STAR) with waveguides etched into the 220 nm device layer of a silicon-on-insulator (SOI) wafer. With the exception of grating couplers and forward biased PIN phase shifters, all waveguides were fully etched. 400 nm wide single mode waveguides used for rings, directional couplers and waveguide bends transition to 2 µm wide multi-mode waveguides in straight waveguide sections in order to reduce propagation losses.

Light is first collected from the MLL via an inverse tapered edge coupler, in which the silicon waveguide is tapered down to 200 nm at the closest point to the interface (the waveguide is terminated ~1.1 µm before the interface to accommodate overlay tolerances). The facet of the chip was slightly angled (8º) in order to reduce back-reflections, and the inverse taper also angled inside the chip with a correction taking refraction of the beam into account. Given a typical mode field diameter (MFD) of 3 µm at the input of the chip, this angle allows reduction of the reflection back into the laser by an additional 3 dB relative to the -14 dB otherwise expected from the interface. Based on previous measurement results, insertion losses between MLL and on-chip waveguide are expected to be 3 dB to 6 dB as the distance between the MLL and the PIC is increased from 1.5 µm to 5 µm [11].

Two rings each serve to selectively couple out one comb line from the main bus waveguide and to drop it to an auxiliary waveguide. Each of these auxiliary waveguides form part of a delay line and are connected to a forward biased PIN phase shifter followed by a Sagnac loop acting as a reflector.

The rings themselves are thermally tuned with titanium nitride (TiN) heaters implemented in the back-end of the chip [30] and were implemented with a radius of 10 µm resulting in an FSR of 8.7 nm that was assessed to be wide enough for adjacent resonances to fall outside of the spectrum of the MLL. They can be thermally tuned at a rate of 70 pm/mA². The loaded quality (Q-)factor of the resonators was measured to be 20,000, corresponding to a ring coupling coefficient of $\kappa^2 = 2.41\%$ (to either waveguide). Based on this and on measured single mode waveguide losses of 5 dB/cm, we expect the insertion losses from bus to drop waveguides to be on the order of 1.2 dB [31]. One of the rings, referred to as ring 2 in the following, features pronounced Mie splitting [32] with 76.9 pm resonance splitting (see Suppl. Mat.).

While we showed nonlinear effects in the rings to be playing a role when applying feedback to a high power DFB [24], the much lower power levels obtained here from single MLL lines exclude such a possibility. Given the finesse of the rings, the characteristics of the laser, and the edge coupler losses, we estimate the power circulating inside the rings to be no more than 5 mW. Moreover, given the rings' full width at half maximum (FWHM) of ~10 GHz, they are not suited for directly shaping the optical spectrum of the dropped comb lines. Rather, they are modeled as simply adding a tunable attenuation and an additional delay to the delay loops, as discussed below.

The Sagnac loops were implemented with 2 by 2 multi-mode interferometers (MMIs), with a 6 µm width and a 43.25 µm length, followed by a waveguide loop. The insertion losses of the MMIs were measured on separate test structures and found to be below 0.5 dB (and below our ability to measure reliably in these test structures). The insertion losses of the phase shifters at 0 V were negligible, however due to the excessively wide intrinsic region width that was chosen, they proved to be very inefficient with thermal effects working against free carrier induced refractive index changes. Consequently, both the tuned feedback phase as well as the feedback strength are determined by the rings in the following experiments.

A number of monitor ports were implemented (Fig. 1(c)). These, as well as the main bus waveguide, were terminated by grating couplers (GCs) used to monitor the transmitted signals.

Based on the device characteristics, the back-reflection from the tunable feedback paths is calculated to be below -10 dB. The cumulative waveguide length separating the chip facet from the Sagnac loops is ~880 µm for both paths, resulting in round trip group delays of ~26 ps in addition to the ~66 ps arising from the rings at resonance. In addition to the back-reflection through the Sagnac loop, ring 2 might also generate some direct reflection due to Mie splitting when tuned into an MLL comb line [33].

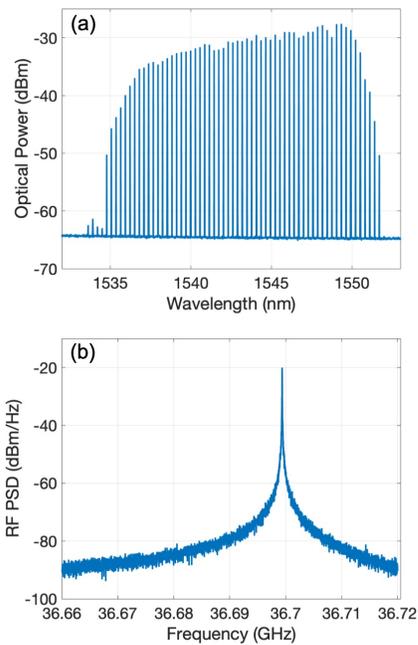

Fig. 2. (a) Optical spectrum of the MLL after transmission through the SiP chip. The 51 pm resolution bandwidth (RBW) of the optical spectrum analyzer is sufficiently large for the entire power of the comb lines to fall within, hence the y-axis is labeled with dBm. (b) Free-running RF spectrum recorded with a RBW of 24 kHz featuring a single RF beat line.

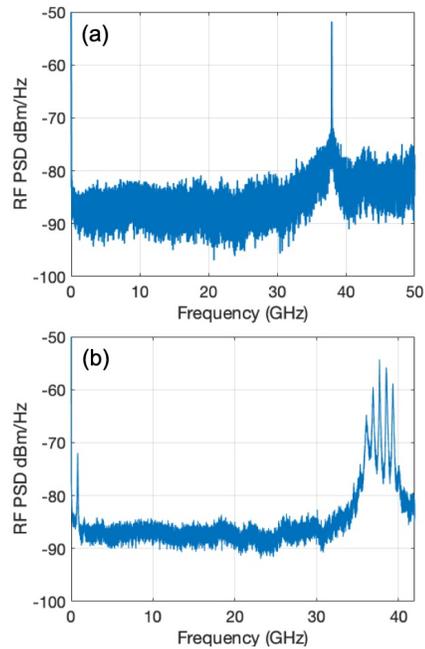

Fig. 3. Wideband RF spectra of an MLL that is nominally identical to the one used to record the other data sets reported in this paper. (a) and (b) were taken at 25C and at 240 mA and 252 mA, respectively. (a) features a single narrow line with a frequency corresponding to the longitudinal mode spacing of the laser. (b) features pronounced relaxation oscillations as well as much wider RF linewidths of ~60 MHz. Data in (a) and (b) was taken with a 3 MHz RBW.

Even though the facet reflection coefficient is estimated to be smaller than the reflection coefficient from the feedback loops, the applied tunable feedback remains weak compared to the broadband parasitic feedback from the chip facet, as a consequence of the former only being applied to two lines and the latter to 40+ lines. This is however partially compensated by the group delay of the tunable feedback being considerably larger, which leads to a stronger effective feedback strength [29],[34]. Due to the proximity of the chip facet to the laser, its main effect is to modify the overall reflectivity at the output of the laser, which may be guarded against by starting with a high enough reflectivity, e.g., by applying an adequate coating. For example, the reflectivity of an uncoated laser facet is on the order of 36%. Given the back-reflection at the silicon PIC facet of -17 dB, the effective reflectivity of the laser's output facet would vary between 25% and 47%, a substantial swing that would for example affect the lasing thresholds of the different lines. Starting with a laser facet reflectivity of 90% on the other hand, the resulting effective reflectivity would remain in the 87% to 92% range, so that its variability would play a much smaller role in regards to thresholds once internal laser losses are considered.

The GCs at the output of the chip are another important source of reflection, in particular the GC connected to the main bus waveguide, since it also reflects back the entire comb. Based on ripples in GC loops, the waveguide-to-waveguide GC back-reflection has been extracted to be on the order of -17 dB. Taking into account edge coupler insertion losses as well as round trip waveguide losses, the resulting feedback can be expected to be below -25 dB. Even though here too the reflection coefficient is significantly smaller than that of the feedback loops, once it is taken into account that the GC reflection is applied to the entire comb, the reflected power from the GC can be seen to be commensurate and even a bit larger than the reflected power from a single tunable feedback loop (applied to a single line). Moreover, here the reflection is associated to a substantial round trip delay of 44 ps, resulting from the 1.5 mm waveguide length between the edge coupler and the GC. Thus, this parasitic back-reflection is also expected to have a substantial effect on laser mode-locking.

Several possibilities exist to reduce parasitic back-reflections in the future: An anti-reflective coating could be added to the input facet or the input facet angle further increased. While low-reflection GCs typically require very fine lithography [35], more recently ultra-low reflection GCs not requiring grating apodization have also been shown [36].

### B. Detailed Description of Semiconductor MLL

The semiconductor MLL is a Q-dash buried ridge stripe (BRS) laser operated as a single section laser [5], i.e., without a saturable absorber. It has a stripe width of 1.5 µm, 6 layers of InAs Q-dashes in an InGaAsP barrier grown on an InP wafer and a length of 1140 µm resulting in an FSR of 36.7 GHz. The current setpoint of the laser, 240 mA, was chosen to yield a stable and narrow RF linewidth at 30ºC, resulting in a 15.01 kHz 3 dB RF linewidth when free running without optical feedback.

Figure 2 shows the ~13 nm wide laser spectrum centered on 1543 nm, as well as the free running electrical spectrum featuring a single RF beat note at 36.7 GHz. The extracted RF linewidth of 15 kHz (see Fig. 6 for a zoomed in spectrum) is three orders of magnitude below the free running optical linewidth, which is by itself a strong indication of mode-locking. To further corroborate mode-locking, we have taken wideband RF spectra (Fig. 3) starting from DC. Spectra with a narrow RF linewidth feature a single, clean peak at a frequency corresponding to the FSR of the longitudinal modes (Fig. 3(a)), which is typically taken as a sign of mode locking [37]-[39]. As a counter example, we also show a spectrum taken at a different drive current, in which the laser is subjected to pronounced relaxation oscillations (Fig. 3(b)), visible in both the isolated peak close to DC at 807 MHz, as well as in the multiple peaks around 37 GHz that are split by a corresponding amount. In this regime, the RF linewidth of individual peaks reaches 60 MHz and is even larger than the optical linewidth under good mode-locking conditions. Data shown in Fig. 3 was taken with a laser nominally identical to the one with which all other data shown in the main part of the paper (i.e., excluding Suppl. Mat.) was taken. Unfortunately, the first laser had been damaged by the time the data shown in Fig. 3 was recorded, however this second laser was from the same wafer and had an identical layout. The data shown in Fig. 3 is thus characteristic for both lasers.

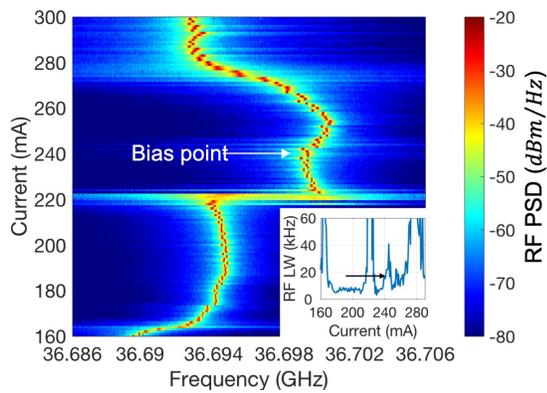

Fig. 4. RF spectra of the MLL as a function of injection current recorded at 30 °C. A sudden jump is apparent in the laser's FSR as the injection current drops below 220 mA. The arrow marks the operating current used in the system characterization. The inset shows the extracted RF linewidths as a function of bias point.

Figure 4 shows the RF spectra of the (1st) MLL as a function of injection current as a color plot. At 220 mA, a jump in the laser's FSR is apparent as it transitions to another collective supermode [40], however a relatively wide range of stable operation is given around the chosen bias point. It should be noted that while the optical feedback is applied to two lines only, RF linewidths reported below are measured from the entire spectrum. Since the PIC applies a weak feedback to an already closed laser cavity, a full spectrum spanning >10 nm is generated as for the stand-alone laser diode.

## III. CHARACTERIZATION RESULTS

Operation of the laser was investigated under optical feedback, i.e., after placing the laser at a few micrometer distance of the PIC's edge coupler and aligning its position with a piezoelectric actuator to maximize the coupling efficiency. Tuning currents in the range 4 mA to 6 mA were independently applied to both rings, shifting their resonances by > 4 laser FSRs around the center of the laser spectrum. Adjacent ring resonances on the edges of the laser spectrum only interact with very weak comb lines, more than 10 dB below the power of the central lines, thus with weak effect on lasing dynamics (see Suppl. Mat. for details). Nonetheless, in view of this it would have been advantageous to slightly increase the ring's FSR, as spurious reflections of these other comb lines might be detrimental due to their uncontrolled phase.

Figure 5 shows a summary of the characterization results with panels (a), (b) and (c) respectively showing the optical power transported to the main output port of the PIC [marked by arrows in Figs. 1(a) and 1(c)], as well as the FSR (as an offset relative to 36.7 GHz) and RF linewidth of the MLL.

Small drops in the transmitted power mark the spectral alignment of a ring with an MLL comb line, which is then dropped and filtered out before reaching the output port (alignment of the resonances with MLL comb lines was also independently verified based on the MLL's spectrum and the rings' tuning curves, see Suppl. Mat.). These ring bias points are marked by white lines in the figure, wherein these occur in pairs for ring 2 as a consequence of the split resonance.

Five broad regions of operation, marked as $R_1$ to $R_5$, are visible in the RF linewidth. They respectively correspond to 3 dB RF linewidths on the order of ~5 kHz, 80 kHz, 20 kHz, several hundred kHz and 3 kHz. While the RF linewidths are better than that of the free running laser in regions $R_1$ and $R_5$, and comparable (if slightly worse) in region $R_3$, in regions $R_2$ and $R_4$ the RF linewidths are severely broadened. These 5 regions are independent of the spectral alignment of the rings, since they are much wider than the regions in which the ring 2 resonance overlays with a comb line. Rather, they are a consequence of the broadband reflections from the PIC to the MLL (including reflections from the PIC facet as well as on-chip devices such as grating couplers), that, depending on their phase, either help or hinder mode-locking, or of temperature drifts in the environment. Slow drifts in the test setup can change the phase of the broadband reflections, for example due to overall heating of the chip and its submount over time. Indeed, in the automated measurement, the ring 1 current sweep is embedded inside the ring 2 sweep, i.e., while ring 1 is rapidly and repeatedly swept between 4 mA and 6 mA, ring 2 is only swept once at a much slower pace. Thus, the ring 2 bias can also be taken as the time axis (21 hours total).

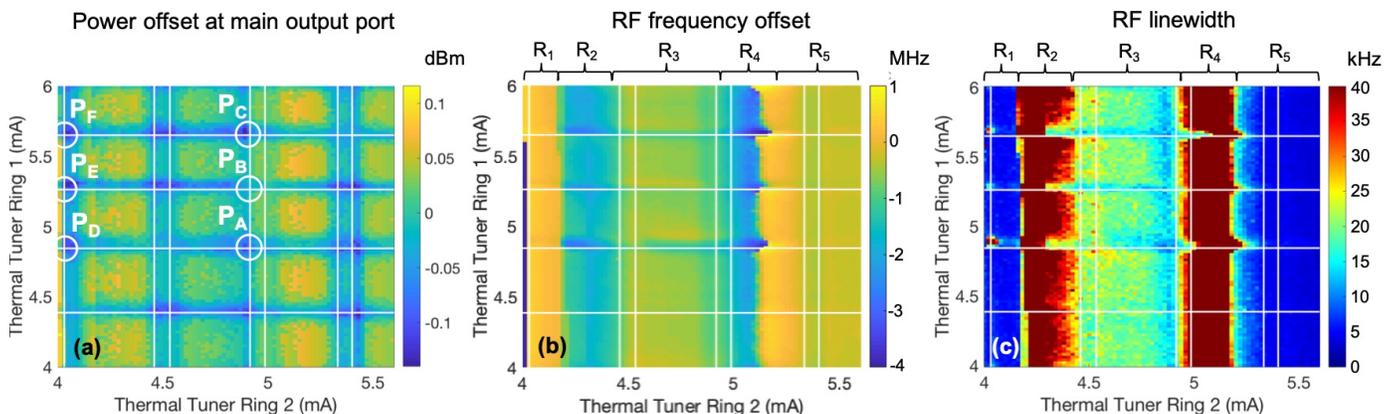

Fig. 5. Characterization of the MLL under feedback from two independently tuned rings. (a) Optical power recorded at the main output port of the PIC [marked by an arrow in Fig. 1(a)]. Color coding shows the power in dBm relative to the average output power. Changes are slight, as at most two lines out of ~40 are filtered out. (b) FSR offset relative to 36.7 GHz and (c) RF linewidth of the MLL. White lines indicate bias points at which a ring resonance is aligned with a comb line. Regions $R_1$ to $R_5$ correspond to a slow drift of the broadband reflection phases. In the measurement sequence, scanning of ring 1 bias is embedded into a single ring 2 scan, so that the ring 2 bias can also be taken as the time axis (21 hours total) during which the laser-to-PIC distance slowly drifted. Data in (c) was recorded with a 16 kHz RBW.

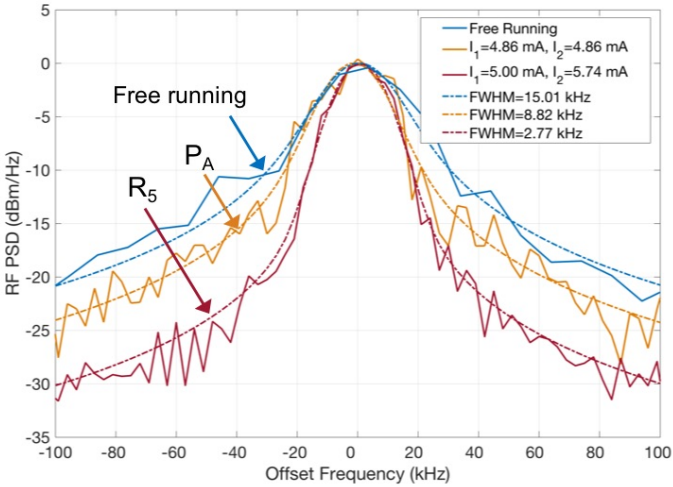

Fig. 6. Recorded RF spectra for the free running laser (blue curve) and under feedback for ring bias points $P_A$ ($I_1 = 4.86$ mA, $I_2 = 4.86$ mA) and a typical spectrum from region $R_5$ ($I_1 = 5.00$ mA, $I_2 = 5.74$ mA). Voigt profile fits are also shown with dash-dotted curves and the linewidth of the underlying Lorentzian, convoluted with the known Gaussian frequency response of the spectrum analyzer, indicated in the legend. Plots have been individually normalized to 0 peak power.

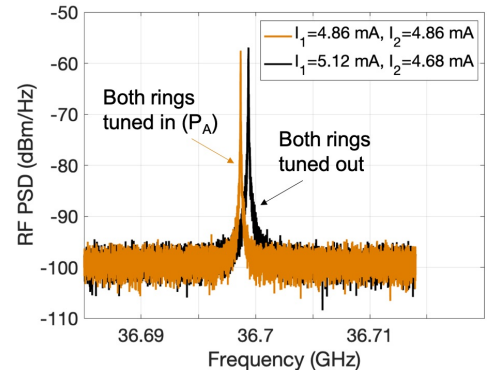

Fig. 7. Wider band RF spectra for the feedback condition $P_A$ ($I_1 = 4.86$ mA, $I_2 = 4.86$ mA) also shown in Fig. 6 (orange), as well as for a different tuning condition in which both rings are tuned out of MLL comb lines ($I_1 = 5.12$ mA, $I_2 = 4.68$ mA). Both spectra feature a unique RF beat note. The slight reduction in the RF peak power at $P_A$ (0.5 dB) is attributed to the fact that two comb lines are filtered out by the PIC, thus reducing the recorded signal.

The good mode-locking regions $R_1$, $R_3$ and $R_5$ feature different laser FSRs respectively offset by ~0 MHz, -0.5 MHz and -0.25 MHz from 36.7 GHz. The FSR remains stable within a given mode-locking region. The poor mode-locking regions $R_2$ and $R_4$ on the other hand correspond to transitions between different stable mode-locking regimes. In these transitions, in which the RF linewidth is considerably broadened up to the MHz range, the MLL FSR is also lowered by about ~2 MHz relative to $R_1$ and $R_5$. As we will see in the following, this gives the opportunity to study the effect of controlled filtered feedback both within a mode-locking region ($R_3$) and in a transition between mode-locking regions ($R_3$ to $R_5$). While the first will result in an improvement by a factor > 2, the second will result in an extension of the stable mode-locking region, thus yielding much higher improvements up to a factor 40.

### A. Filtered feedback within stable region $R_3$

It is apparent that frequency selective feedback can modify the RF linewidth when a ring resonance overlays with an MLL comb line. Here too, the linewidth is reduced or enhanced depending on the feedback phase, but since the latter is also changed by the exact ring bias point, it can be fine-tuned with the ring provided the feedback phase at resonance is close enough to target. Since the phase of the light transmitted between the bus and drop ports is evaluated to change by $0.85\pi$ as the ring resonance is tuned through the MLL line, a total round trip phase tuning range of $1.7\pi$ can be reached this way (in the most general case, the phase shifters would ensure that a constructive interference condition can always be reached independently of the ring bias point and thus of the chosen feedback strength). E.g., in region $R_3$ a crisscross pattern of deeper blue regions (in which the RF linewidth is improved) is seen in Fig. 5(c), corresponding to ring resonances overlaying with MLL comb lines (as shown by the white lines). The RF linewidth is improved from ~20 kHz down to 8.82 kHz at the point marked by $P_A$ in Fig. 5(a). The RF linewidth minima can be seen to occur slightly to the left of the ring 2 resonances, with the lower bias currents corresponding to a detuning of about one half of the resonance's FWHM (as determined from the thermal tuning curves of the rings and the recorded optical spectrum of the MLL). At points $P_D$, $P_E$ and $P_F$ on the other hand, the RF linewidth is increased from the surrounding ~5 kHz to respectively 133 kHz, 14.5 kHz and 26.9 kHz as a consequence of the destructive interference condition. In addition to the ring's thermal tuning curves, the drop in the optical power levels transmitted to the main output GC of the chip, as seen in Fig. 5(a), provides independent confirmation that points $P_A$ to $P_F$ coincide with both rings being tuned into MLL comb lines.

RF linewidths were measured with a resolution bandwidth (RBW) of 16 kHz, to avoid prohibitively long measurement times. Measured spectra have been fitted to a Voigt profile resulting from the convolution of a Lorentzian with the known Gaussian frequency response of the spectrum analyzer, with good fitting accuracy and stability (Suppl. Mat.). Numbers reported here correspond to the FWHM of the underlying Lorentzian linewidth. Figure 6 shows exemplary spectra in a small range around the RF beat note together with the Voigt profile fits. Clear improvements relative to the free running spectrum can be seen at both the tuning point $P_A$ and in the exemplary spectrum from feedback range $R_5$.

Figure 7 shows the spectra for tuning point $P_A$ (both rings tuned into separate MLL comb lines) as well as for another exemplary tuning point within range $R_3$ in which both rings are detuned. It is apparent that both spectra feature a unique RF beat note and that their maximum power is of comparable magnitude, with a slight decrease in the case of $P_A$ (0.5 dB) attributed to the signal reduction associated to two comb lines being filtered out. In fact, this results in 3 beat note contributions to be removed out of ~40 (the beat tone between the two adjacent, filtered lines, as well as the two beat notes with their nearest neighbors), roughly consistent with the observed signal strength reduction. This comparison between RF spectra with and without ring induced feedback is included here to show that the application of such feedback does not result in a collapse of the RF peak power, as would be expected e.g. if a regime with supermode competition (respectively multiple intra-cavity pulse trains in saturable absorber based MLLs [20]), were triggered. Please note that while the peak RF power levels in Figs. 2(b), 3 and 7 are not directly comparable due to different test conditions (different RBW, coupling directly to fiber or through chip etc.), the datasets shown in Figs. 7 and 9(c) were taken closely apart in the exact same test conditions and are thus directly comparable.

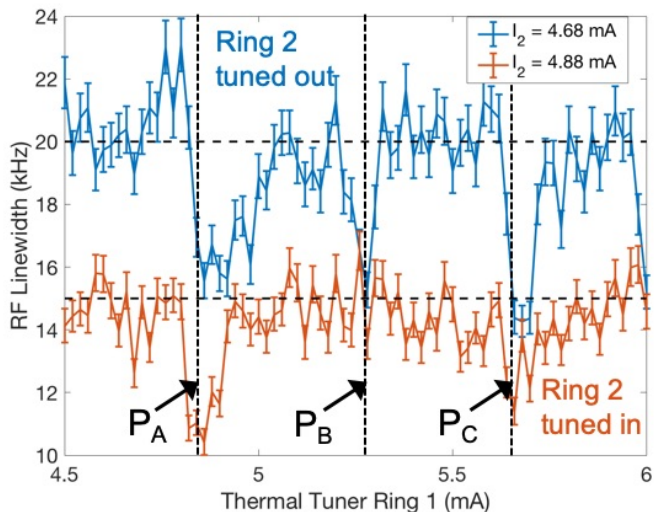

Fig. 8. RF linewidths of the MLL (underlying Lorentzian linewidth as extracted with a Voigt profile fit with a fixed Gaussian component), under feedback as a function of the tuning current applied to ring 1. The blue and red curves correspond to different ring 2 bias points, with one tuning curve passing through $P_A$, $P_B$ and $P_C$ (red curve corresponding to ring 2 tuned in at $I_2 = 4.88$ mA) and the other passing to their left in Fig. 5 (with ring 2 tuned out to $I_2 = 4.68$ mA). The error bars show the $3\sigma$ confidence intervals defined as the ranges in which the residual sum of squares increases by 9 times the estimated variance of a single data point in the recorded spectra. Vertical dashed lines show the ring 1 bias points corresponding to points $P_A$, $P_B$ and $P_C$ as determined from Fig. 5(a) and from ring 1's tuning curve (Suppl. Mat.).

Figure 8 shows the extracted RF linewidths as the tuning current of ring 1 is swept but the tuning current of ring 2 is maintained either around 4.68 mA, blue curve labeled as "ring 2 tuned out", or 4.88 mA, red curve labeled as "ring 2 tuned in" (for better statistical significance each curve corresponds to an average of -0.02 mA, 0 mA and +0.02 mA from the nominal ring 2 bias current). In the blue curve, ring 2 is maintained detuned relative to the MLL comb lines, while in the red curve ring 2 remains tuned to an MLL comb line throughout. In both, ring 1 moves in and out of alignment with the MLL lines. The red tuning curve connects points $P_A$, $P_B$ and $P_C$, while the blue runs to its left in Fig. 5.

It is apparent that the latter features a reduced RF linewidth at three tuning points, whenever ring 1 overlaps with an MLL comb line. The baseline of the red curve is at that same reduced level, as a consequence of the other ring, ring 2, always being aligned with a comb line. When both rings are aligned to separate comb lines (points $P_A$ and $P_C$, as labeled), further enhancement is visible as a consequence of the collective action of both feedbacks. When both rings are aligned to the same comb line, no significant further improvement is observed ($P_B$). Since very little of that comb line reaches the downstream ring, no additional feedback is created by the latter. The improved feedback under the collective effect of both rings at points $P_A$ and $P_C$ might be explained by the feedback qualitatively containing more information, as the beat tone between the two comb lines also provides feedback in respect to the RF phase, as opposed to only the optical phase as in the case of a single line. It is however also possible that it is simply a consequence of the feedback being quantitatively stronger, as the power from two lines is being sent back.

As the 20 kHz linewidth in region $R_3$ has been reduced to 8.82 kHz, the best linewidth at $P_A$, it is apparent that a linewidth narrowing with a factor better than 2 has been obtained. This is a substantial improvement and beats even the linewidth of the free running laser by a factor 1.7, even though the baseline linewidth in region $R_3$ starts out slightly worse. Thus, detrimental operating conditions seen in region $R_3$ are more than compensated. However, it is still not as good as in regions $R_1$ and $R_5$ in which even smaller linewidths have been obtained, respectively on the order of 5 kHz and 3 kHz. The improvement in regions $R_1$ and $R_5$ may be due to the broadband reflection itself being favorable and contributing to linewidth reduction. As an example, Fig. 6 also shows a typical spectrum obtained in $R_5$, with an extracted linewidth of 2.77 kHz corresponding to an improvement better than a factor 5 relative to the free running laser.

### B. Filtered feedback within instability region – transition from $R_3$ to $R_5$

In the following, we will see that much more substantial improvements have been achieved in regimes in which the RF linewidth of the MLL starts out as being very significantly deteriorated, to the point of recovering the free running RF linewidth (~15 kHz) in a wide range of highly unfavorable operating conditions in which phase locking between the modes had been almost completely destroyed (up to 572 kHz RF linewidth without stabilization). In the most extreme case, an RF linewidth of 1.5 MHz is narrowed back down to 40 kHz, which remains sufficient for the initially targeted direct detection WDM link application [41].

Indeed, an important question is how well the stabilization scheme works under very adverse broadband reflections severely broadening the RF linewidth beyond the free running condition, for example due to an ill-controlled laser to chip-facet distance in a flip-chip attachment process [11] or fabrication/temperature variations modifying the on-chip optical path length to a reflective device, as mitigation of such adverse conditions is one of the objectives of this work. In regions $R_2$ and $R_4$, that correspond to a significant degradation of the RF linewidth, ring resonator 2 remained tuned out of the MLL lines, so that dual ring feedback could not be analyzed

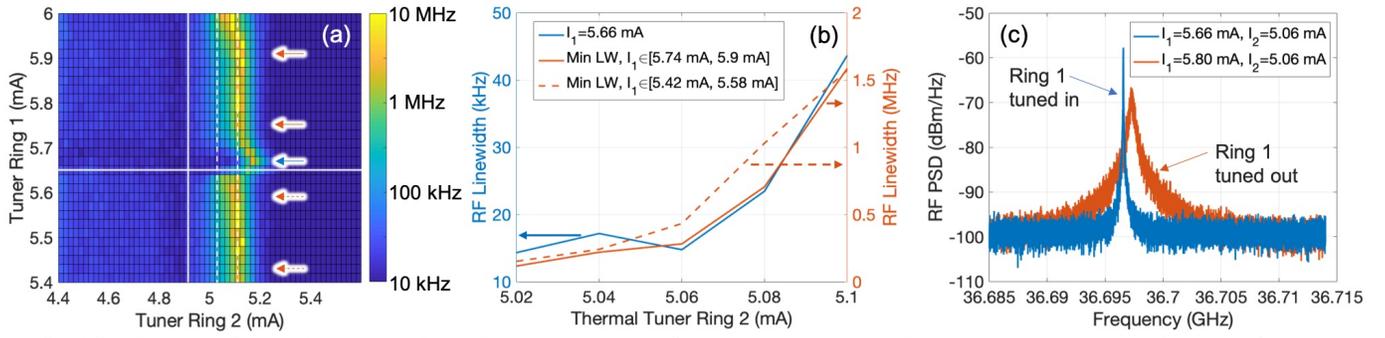

Fig. 9. (a) Detailed view of Fig. 5(c) in a range of overall poor mode-locking ($R_4$). In order to better visualize the large dynamic range of observed RF linewidths, the color scale maps to the decimal logarithm of the linewidth. As in Fig. 5, the continuous white lines represent tuning currents at which ring resonances align with a comb line. Dashed lines indicate the ring 2 tuning current range and arrows the ring 1 tuning currents for which the RF linewidths are better plotted in (b). The blue (left axis), continuous red (right axis) and dashed red (right axis) curves respectively correspond to (i) a ring 1 tuning current of 5.66 mA (corresponding to ring 1 being tuned into an MLL line), (ii) the minimum RF linewidth in the 5.74 mA to 5.9 mA ring 1 current range and (iii) the minimum RF linewidth in the 5.42 mA to 5.58 mA ring 1 current range. (c) shows exemplary RF spectra for the same ring 2 tuning current, but different ring 1 currents of 5.66 mA (blue curve) and 5.80 mA (red curve) corresponding respectively to ring 1 being tuned in and out of an MLL comb line. These also correspond to points of the blue and red curves in (b) with common abscissa $I_2$=5.06 mA. As in previous datasets, a single RF beat note is visible.

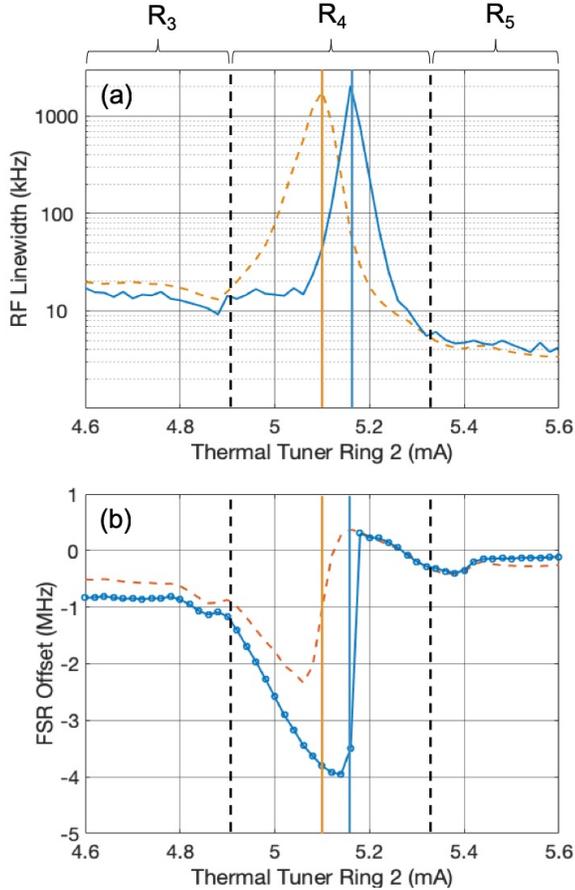

Fig. 10. (a) RF linewidth with and without feedback from ring 1, wherein the solid curve corresponds to ring 1 feedback ($I_1$ = 5.66 mA as in Fig. 9) and the dashed orange curve corresponds to the median of the recorded RF linewidth without ring 1 feedback. As previously, the ring 2 bias point is an indirect metric for elapsed time. (b) FSR with (solid blue curve) and without ring 1 feedback (dashed orange curve) in the same time window. Here too, the dashed curve is the median value without feedback. The y-axis indicates the offset of the FSR from 36.7 GHz. Dashed vertical lines show the boundaries between regions $R_3$, $R_4$ and $R_5$ in both graphs. Solid vertical lines highlight the coincidence of the point of worst linewidth with the center of the supermode transition. The data for the case of single ring feedback is also plotted with markers to show the steepness of the transition, which completely fits between two data points.

there. However, the effect of feedback from a single ring (ring 1) could be analyzed.

As seen in Fig. 9(a), feedback from ring 1 allows a substantial reduction of the RF linewidth in region $R_4$, in which it is otherwise significantly deteriorated. As further seen in Fig. 9(b), the obtained linewidth under feedback (blue curve) is about a factor 20 to 40 better than the linewidth without ring 1 feedback (continuous and dashed red curves). As previously explained, the ring 2 tuning current shown on the x-axis is a proxy for measurement time, during which operating conditions become progressively worse. The dashed and continuous red curves show the RF linewidths immediately before and after ring 1 feedback was applied, to rule out drift being the cause of the improvement. At the point $I_2$ = 5.06 mA, the linewidth is for example improved from 280-430 kHz to 14.8 kHz as ring 1 feedback is applied. Corresponding RF spectra are shown in Fig. 9(c).

In the worst case shown in Fig. 9(b) corresponding to $I_2$ = 5.1 mA, in which the parasitic broadband feedback or thermally induced transition between supermodes broaden the RF linewidth to 1.5 MHz, tuned feedback with one ring results in an RF linewidth of 40 kHz. While this remains worse than the performance of the solitary laser (15 kHz), it is noteworthy that the improvement is almost a factor 40. Moreover, at 1.5 MHz, the RF linewidth is relatively close to the optical linewidth of the laser and phase locking between the modes is thus seen to be extremely weak. Importantly, orders of magnitude better phase locking can still be recovered in this regime by means of controlled feedback of a single line.

In order to obtain a better understanding of the functioning and limitations of the stabilization scheme, Fig. 10(a) compares the RF linewidth with and without feedback from ring 1 in a wider time window (represented by the ring 2 current), wherein the solid curve corresponds to $I_1$ = 5.66 mA as previously (ring 1 tuned in) and the dashed orange curve corresponds to the median of the recorded RF linewidth without ring 1 feedback (the baseline). Figure 10(b) shows the FSR (offset relative to 36.7 GHz) with and without feedback in the same time window.

Looking first at the FSR without ring 1 feedback, it is apparent that the FSR transitions between two supermodes as a consequence of the changing parasitic feedback (the transition occurs between ring 2 currents 4.9 mA and 5.2 mA). At the onset of the transition, the FSR of the initial supermode is first pulled down, after which the system rapidly switches to the other supermode, resulting in an S-shaped transition. The RF linewidth starts deteriorating at the onset of the transition and reaches its worst value halfway through the supermode switch.

Under ring 1 feedback, it is apparent that the initial supermode, while also being pulled to a lower FSR, remains stable in a larger region (as witnessed by the low RF linewidth), delaying the supermode switch that then occurs much more abruptly. The RF linewidth, as stabilized by feedback from ring 1, also remains relatively immune to the changing broadband feedback, maintaining a linewidth of ~15 kHz, until the baseline RF linewidth reaches 572 kHz, after which it relatively quickly deteriorates. Interestingly, 15 kHz is also the linewidth under single ring stabilization obtained in the stable regions of $R_3$ far from the supermode transition (Section III.A), so that it appears the exact same performance is recovered here. The worst case RF linewidth is reached here too in the middle of the bifurcation. Moreover, the region of poor mode-locking is smaller, due to the sharp transition.

It is apparent that the ring 1 feedback extends the region of stability of the first supermode, while it shrinks the range of stability of the second supermode (possibly as a consequence of a bistability given the abruptness of the transition). After the switch, the RF linewidth is also worsened rather than improved by ring 1 feedback. This may be due to the dialed in feedback phase, which could also explain why prior to the switch the FSR of the laser under tuned feedback is below that of the laser without tuned feedback, while after the switch the opposite is the case. Indeed, it could be indicative of the resonator induced feedback resulting in frequency pulling on the fed back comb line in opposite directions before and after the switch, also affecting the overall FSR of the laser in opposite ways via the four wave mixing nonlinearities at the origin of mode-locking in single section MLLs [5]. Similarly, inspecting the FSR in the ring tuning regions reported in section III.A reveals there too RF linewidth enhancement was accompanied by a reduction of the FSR, by about ~0.62 MHz after tuning in the first ring (in the vicinity of tuning point $P_A$), followed by an additional ~0.23 MHz after tuning in the second ring (at point $P_A$). This is not surprising considering the above, since point $P_A$ is also located in region $R_3$, so that the same linewidth reduction mechanism is expected. Thus it is too early to generalize this trend as there could be a number of causes in the laser dynamics. However, reduction of the RF linewidth as the FSR is shifted down may be analogous to the optical linewidth of a single wavelength laser being reduced when its emission frequency is pulled down, as predicted by the Lang-Kobayashi equations [34],[42]. A rigorous interpretation will require detailed modeling.

A study revealing more systematically the limits and capabilities of this stabilization scheme for different mode locking conditions would require improved long term mechanical stability, e.g. with a flip-chipped MLL, as well as functional phase shifters to deconvolute the feedback phase from the feedback strength. Moreover, while it has been verified that phase locking between longitudinal laser modes is maintained and improved, further investigation with autocorrelation traces (after dispersion compensation) would be beneficial to verify that the method utilized here does not deleteriously impact pulse shape in the time domain.

## IV. Discussion and Benchmarking

While not allowing the laser to be fully immune to other detrimental effects (which would not have been expected), reasonable operating conditions could be obtained, in particular in view of the less stringent applications such as suppressing mode partition noise for multi-channel direct detection communication systems, even starting from severely deteriorated mode-locking. In a complete system combining an MLL stabilization scheme with modulator arrays, e.g. in the form of cascaded resonant ring modulators, insertion losses associated to the latter as well as extinction of unused comb lines by channel selection filters [43] could further contribute to reducing parasitic broadband feedback.

In order to compare the results reported here to others from the literature, a figure of merit (FOM) is required that takes both the RF linewidth and the repetition frequency into account, as achievable RF linewidths typically become worse with increasing repetition rate [44]. In RF electronics, microwave sources realized by upconverting a reference frequency generated by a quartz oscillator typically have a phase noise that degrades by 6 dB for every doubling of the signal frequency [45], corresponding in a quadrupling of the oscillator linewidth [17]. A similar scaling law would result from pulse picking (e.g., taking out every second pulse to halve the repetition rate), so that one could first expect this sizing law to also apply to semiconductor MLLs (doubling the cavity length could be seen as being analogous, as pulses are then extracted at half the rate, as if every second pulse had been picked out). In [44], however, the RF linewidth of single-section semiconductor MLLs of varying length was rather found to scale almost linearly with the FSR, so that here the ratio of FSR to RF linewidth is taken as FOM. This choice of FOM results in a more conservative comparison of our results to previous, lower FSR results reported in the literature, as previously reported RF linewidths are then scaled up less aggressively to obtain their equivalent value at 36.7 GHz.

TABLE I
RF Linewidths of Semiconductor MLLs with Silicon Delay Lines

| Ref. | Method | FSR | 3 dB RF Linewidth | Rescaled 3 dB RF Linewidth |
|---|---|---|---|---|
| [18] | Intra-cavity delay line | 2 GHz | 14 kHz | 257 kHz |
| [18] | Intra-cavity delay line & intra-cavity filter | 20 GHz | 52 kHz | 95 kHz |
| [21] | Broadband external feedback | 17.4 GHz | 29 kHz | 61 kHz |
| [16] | Intra-cavity delay line | 1 GHz | 0.3 kHz[1] | 11 kHz |
| [22] | Broadband external feedback | 19.2 GHz | 6 kHz | 11 kHz |
| This Work | Best operating conditions | 36.7 GHz | 2.8 kHz[2] | 2.8 kHz |
| | Dual comb feedback applied to a 20 kHz baseline | 36.7 GHz | 8.8 kHz[2] | 8.8 kHz |
| | Single comb feedback applied to a 572 kHz baseline | 36.7 GHz | 15 kHz[2] | 15 kHz |

[1]Extrapolated from the 10 dB linewidth assuming a Lorentzian shape.
[2]These correspond to the Lorentzian component of a Voigt fit.
Results are listed in chronological order.

Table 1 summarizes the performance of previously reported semiconductor MLLs stabilized with silicon delay lines. The 10 dB RF linewidth reported in [16] was converted into a 3 dB linewidth assuming a Lorentzian shape. The column "Rescaled 3 dB RF Linewidth" reports the linewidth that would result in the same FOM at the 36.7 GHz comb spacing reported here, i.e., the RF linewidth is rescaled proportionally to the FSR. It can be seen that the linewidths under favorable broadband reflection (2.77 kHz) and under slightly detrimental broadband reflection and dual ring stabilization (point $P_A$, 8.82 kHz) compare favorably to the state of the art. Even in the presence of highly detrimental broadband feedback (with baseline RF linewidths up to 572 kHz), reduction of the RF linewidth to ~15 kHz remains in the range of the better results.

For fair comparison, it should be pointed out that gain materials utilized in [16],[18],[21],[22] were quantum well based, while the gain material used here is Q-dash based, which may be favorable to lower linewidths [46]. Even lower RF linewidths than reported here have been obtained with single section semiconductor MLLs using Q-dash gain material in combination with a highly delocalized optical mode [44]. RF linewidths as low as 300 Hz have been obtained with long cavity devices. The shortest devices, with a cavity length of 2 mm and an FSR of ~21 GHz, are reported to have an RF linewidth of ~1.5 kHz. Extrapolating the dependency of the RF linewidth on cavity size observed in [44], one would expect a linewidth of ~2.6 kHz at an FSR of 36.7 GHz, comparable to the best results obtained here under favorable operating conditions. However, due to the reduced modal confinement and smaller differential gain, thresholds are also significantly increased, reaching 354 mA for the 2 mm long device (corresponding to 3.6X the threshold current density of baseline devices without mode delocalization). As a consequence of the smaller differential gain, further shrinking of these devices and corresponding scaling to larger FSRs does not appear entirely straightforward. As the work reported here was targeting the demonstration of both large FSRs and narrow linewidths, we purposefully opted for a gain material without reduced modal confinement in order to maintain high differential gain.

One may remark that while the obtained RF linewidth improvements are adequate for the main application pursued here – the utilization of single section semiconductor MLLs as multi-carrier sources for direct detection communication systems – others, in particular in the field of metrology, would require much more regular pulse trains. In order to achieve substantially lower RF phase noise operation, longer low loss delay lines (resonant or linear) would be desirable. Such could e.g. be PIC platforms with low loss, silicon nitride (SiN) waveguides. Platforms with high confinement (and thus densely routable) waveguides have been shown with Q-factors as high as a couple of millions [47]. Even higher Q-factors of 80 million have been obtained in low confinement platforms [48]. Corresponding waveguide losses on the order of ~1 dB/m and ~0.1 dB/m are respectively 2 and 3 orders of magnitude better than single mode silicon waveguides. Delays equivalent to a few meters of fiber [3] can be obtained in such platforms, giving a prospect of much more substantial RF linewidth improvements.

We believe that scaling the length of on-chip delay lines to the order of a few meters would be easier to realize in the architecture proposed here than with external broadband feedback or intra-cavity filters, as shown in [18],[21],[22], since dispersion related issues become more pronounced for the latter as delay lengths are increased.

In order to obtain the right feedback phase for all the comb lines in case of broadband reflection, the external group delay should be close to an integer of the cavity roundtrip time [20],[49], requiring in the ideal case dispersionless waveguides with a perfectly targeted group index. To maintain the phase error within $\pm\pi/4$ for all the comb lines spanning a spectrum of width $\Delta\lambda$ with a round trip delay length $L$, the fabrication error of the group index $\Delta n_g$ and the waveguide chromatic dispersion parameter $D$ have to be bounded by (derived in Suppl. Mat.)

$$|\Delta n_g| \leq \frac{\lambda_0^2}{4L\Delta\lambda} \tag{1}$$

$$|D| \leq \frac{\lambda_0^2}{L c_0 \Delta\lambda^2} \tag{2}$$

wherein $\lambda_0$ is the free-space center wavelength of the laser spectrum and $c_0$ is the speed of light in vacuum. For a 6 meter delay line as in [3] and a 10 nm laser spectrum, this results in $\Delta n_g \leq 10^{-5}$ and $D \leq 13 \; ps/(nm \cdot km)$. Tellingly, the upper bound for the dispersion is close to that of standard single mode fiber. Since the typical sensitivity of the group index of silicon photonics waveguides on their width is on the order of $2 \cdot 10^{-3}$ per nm, a level of control within $10^{-5}$ does not seem feasible.

Similar constraints apply to a laser with an intra-cavity delay line combined with an intracavity filter. Such a filter can be implemented as a ring resonator whose FSR is a multiple of that of the unfiltered laser cavity [18], with the objective of increasing the repetition rate of the laser. The longer the intra-cavity delay line, the smaller the unfiltered FSR of the laser, constraining the linewidth of the intra-cavity filter. As the required filter linewidth diminishes, the spectral alignment of the filter's resonances also has to be more precise, imposing a tight control on the group index of the waveguide out of which the filter is formed. Tellingly, if we take as a criterion that the filter's resonances at the edges of the MLL spectrum should be detuned by no more than an eighth of an FSR, we obtain the exact same constrains as reported above for external broadband feedback [Eqs. (1) and (2)]. In [3], the required control of the group index as given by (1) was circumvented, as the static filter was replaced by an actively switched attenuator in the form of an embedded Mach-Zehnder modulator.

These constraints do not apply to the scheme described here, as phase errors at filtered MLL lines can be compensated with simple phase shifters as opposed to broadband dispersion control. Of course, scaling up delay lines by orders of magnitude will not come without other difficulties. For example, as delay lines become very long and the resulting Fabry-Perot resonances closely spaced, instabilities may appear in form of ripples in the phase noise spectrum of the laser. However, such spurious Fabry-Perot resonances may be filtered out by the ring resonators, or instabilities reduced by providing feedback over several length scales. A comprehensive analysis of the difficulties encountered in aggressively scaling up the delay lines and the realization of such a system remains the object of future work.

## V. CONCLUSIONS

In conclusion, we have shown improvement of the RF linewidth of a single section semiconductor MLL by feedback from a SiP PIC beyond the level obtained from the free running laser even in the presence of adverse operating conditions. RF linewidths as low as 2.77 kHz have been obtained in the best case. Even under adverse conditions with broadband parasitic reflections broadening the RF linewidth to 1.5 MHz, a 40 kHz RF linewidth, adequate for the targeted SiP direct detection transceiver application, could be recovered. Further improvements may require longer, low loss delay lines that could be implemented in a low loss SiN waveguide platform as well as tighter control of the feedback conditions, e.g. by a flip-chip mounted MLL on a rigid PIC, to avoid microphony and drift.

# Semiconductor laser mode locking stabilization with optical feedback from a silicon PIC

## *Supplementary Materials*

Johannes Hauck, Andrea Zazzi, Alexandre Garreau, François Lelarge,
Alvaro Moscoso-Mártir, Florian Merget, Jeremy Witzens, *Sr. Member, IEEE*

## I. Extraction of RF Linewidths

RF linewidths were measured with the resolution bandwidth (RBW) of the electrical spectrum analyzer set to 16 kHz with Gaussian filtering. As a consequence, the recorded spectra are convoluted with the frequency response of the spectrum analyzer. Assuming the spectra to be initially Lorentzian shaped, the end result would be a Voigt profile with a known Gaussian content. The goal of this section is to verify the adequacy of the Voigt profile fitting for extraction of the underlying RF linewidths.

Fits are made by minimizing the residual sum of squares (RSS) obtained by comparing the data and the model on a dBm scale with data points spaced between -400 kHz and +400 kHz from the center of the spectrum on a 3 kHz grid. Lorentzian linewidth, the spectrum's center frequency and the maximum power are left as free parameters, wherein only tiny adjustments are made to the latter two since they are straightforwardly determined from the data beforehand. The RBW of the spectrum analyzer is fixed during RSS minimization. For the two spectra taken under optical feedback shown in Fig. 6 of the main text, with extracted RF linewidths of respectively 8.82 kHz and 2.77 kHz, Fig. SM1 shows the RSS and the extracted underlying linewidths as a function of the assumed Gaussian filter bandwidth. It is apparent that the RSS is minimized between 14 kHz and 15 kHz for both spectra, very close to the 16 kHz expected from the spectrum analyzer settings. Figure SM1(b) shows the sensitivity of the extracted Lorentzian linewidth on the assumed Gaussian RBW. In both cases, minimizing the RSS while leaving the Gaussian RBW freely adjustable would have resulted in even smaller extracted Lorentzian linewidths, 2.56 kHz instead of 2.77 kHz and 8.79 kHz instead of 8.82 kHz. Moreover, these differences in fitting results are quite small. From Fig. 6 (main text) it is apparent that the fits follow the data closely in both the center region dominated by the Gaussian filter shape and on the sides dominated by the skirts of the Lorentzian transfer function.

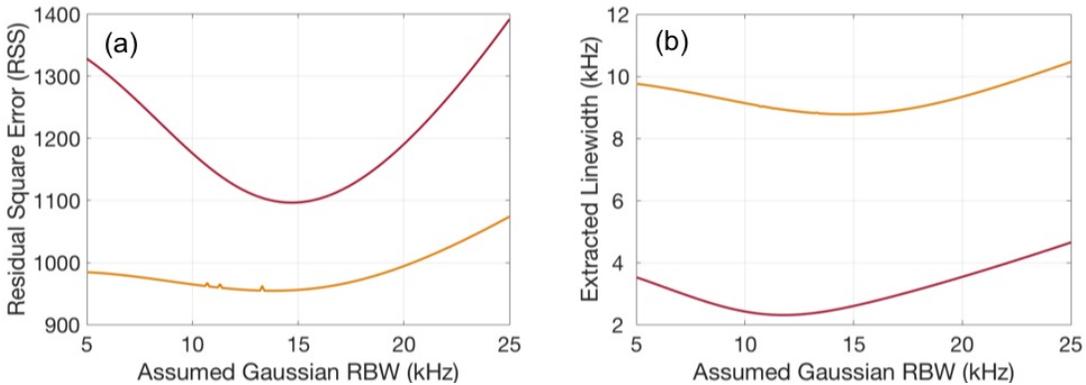

**Fig. SM1. (a) RSS and (b) extracted Lorentzian linewidth for the two spectra under optical feedback shown in Fig. 6 of the main text plotted as a function of the assumed RBW. The same color coding is used as in Fig. 6, i.e., the red and orange curves respectively correspond to the spectra from $R_5$ and $P_\lambda$.**

As a further benchmark of the measurement methodology, we compared a direct measurement of the free running RF linewidth of an MLL nominally identical to the one used for the measurements reported in the main text, as obtained with a relatively small RBW of 1 kHz, to the linewidth extracted from a spectrum recorded with a RBW of 24 kHz with the methodology described above. Unfortunately the laser used for the experiments described in the main text was no longer available due to an electro-static discharge (ESD) event, however this serves to ascertain the soundness of the methodology. Moreover, since the dataset shown in the main text corresponds to 10201 spectra, taking the data directly with a low RBW would have taken a prohibitive amount of time (as is, it already took 21 hours).

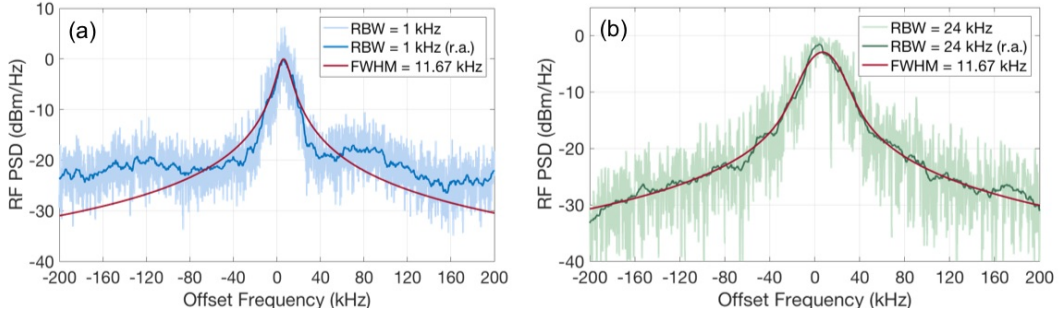

**Fig. SM2. Comparison of RF linewidth measurements (free running MLL) taken with (a) a 1 kHz RBW and (b) a 24 kHz RBW. The 24 kHz RBW spectrum is fitted to a Voigt profile with a fixed Gaussian part and the underlying Lorentzian linewidth extracted (11.67 kHz). The Lorentzian is overlaid over the 1 kHz RBW spectrum shown in (a). The measured PSD are shown both as raw data as well as smoothed with a running average (r.a.). The Lorentzian FWHM extracted from the 24 kHz RBW data appears to fit the 1 kHz RBW data quite well.**

## II. MIE SPLITTING OF RING 2

Via the monitor ports shown in Fig. 1 of the main text, we were able to measure the spectra of the light dropped through the rings to their drop ports as a function of the tuning current (Fig. SM3). The expected parabolic dependence of the resonance wavelength on the current sent through the thermal tuner is clearly visible (the dissipated power is proportional to the square of the current). Both ring 1 and ring 2 feature periodic dips in the transmitted power, exemplarily marked by dashed circles, that are attributed to Fabry-Perot resonances occurring due to spurious reflections inside the PIC. Importantly, it can be seen that the rate at which the center wavelength of these dips varies is different from the rate at which the ring resonance moves (the change of the former being attributed to the variable phase offset resulting from the ring as it is being tuned). In addition, ring 2 features a marked split resonance, exemplarily marked by a continuous circle, wherein both resonances are tuned with the exact same rate. Figure SM4 shows an exemplary transmission spectrum to a monitor port through ring 2. It features the characteristic shape of a split resonance, with a 76.9 pm splitting.

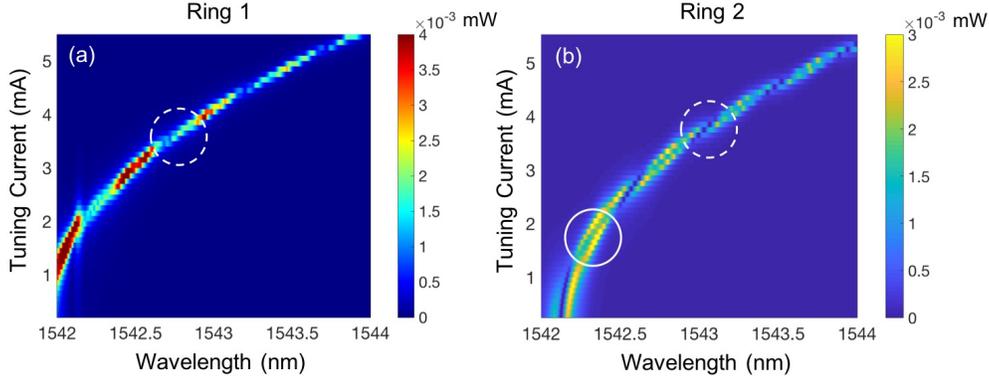

**Fig. SM3. Power transmitted from the main PIC input port (usually connected to the MLL, replaced here by a lensed fiber connected to a tunable laser) through either ring 1 (a) or ring 2 (b) to a corresponding monitor port tapped from the drop waveguide, plotted as a function of wavelength and tuning current. Both ring 1 and ring 2 transmission spectra feature periodic drops in the transfer function (dashed circles) attributed to Fabry-Perot resonances occurring elsewhere in the PIC. In addition, ring 2 features pronounced resonance splitting.**

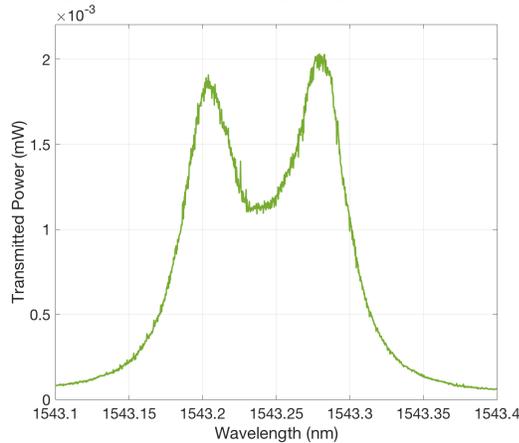

**Fig. SM4. Exemplary transmission spectrum to the drop waveguide of ring 2 as monitored through a monitor tap. The characteristic spectrum of split resonances can be seen, with a resonance splitting of 76.9 nm.**

## III. Thermal Tuning of Rings and Overlay with MLL Spectrum

Figure SM5 shows transmission data recorded between monitor ports connected to the drop side of the rings and the main output port of the chip as a function of tuning current. This data served to confirm the free spectral range (FSR) to be 8.7 nm and to extract the thermal tuning efficiency of the rings.

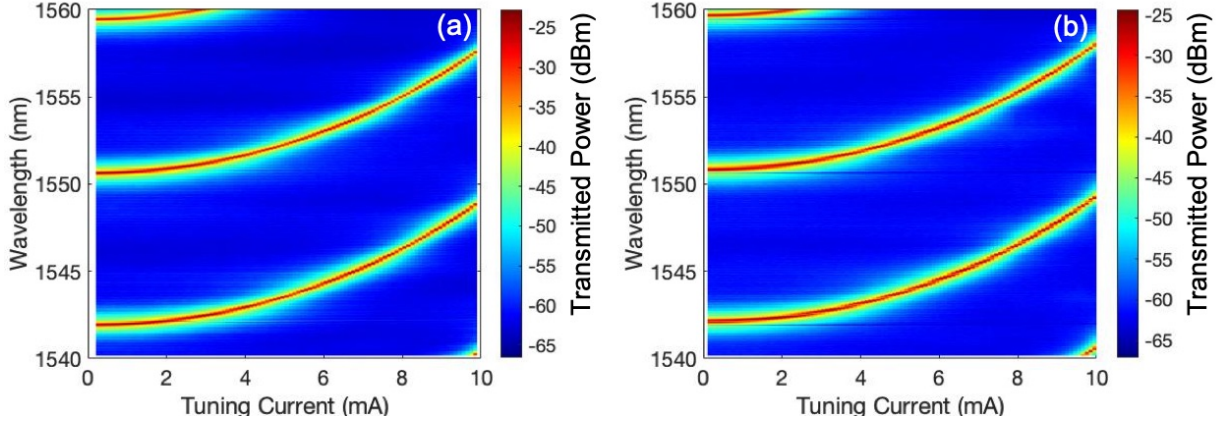

**Fig. SM5. Transmission spectrum between ring specific monitor port and main output of the SiP chip as a function of tuning current, revealing the thermal tuning of the ring resonances. (a) Ring 1 and (b) Ring 2.**

From this data, the resonant wavelengths of rings 1 and 2 were extracted as

$$\lambda_1[nm] = 1541.999 - 0.0644 \cdot I_1[mA] + 0.07722 \cdot I_1[mA]^2$$

$$\lambda_2[nm] = 1542.192 - 0.0586 \cdot I_2[mA] + 0.07716 \cdot I_2[mA]^2$$

wherein it should be noted that the 70 pm/mA² tuning efficiency reported in the main part of the paper was obtained by forcing the small linear terms to zero and making a quadratic fit. These equations were then used to visualize the overlay of ring resonances with the MLL spectrum (as taken from the solitary laser) for the ring tuning points P$_A$, P$_B$ and P$_C$. This overlay can be seen in Fig. SM6. Blue dashed lines indicate the resonant wavelength of ring 1 for the three tuning points. The bias of ring 2 is the same for all three tuning points and corresponds to the central of the three wavelengths, labeled as P$_B$.

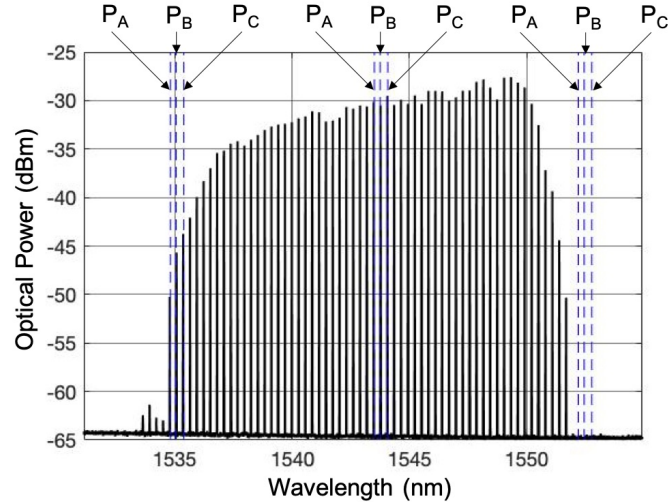

**Fig. SM6. Overlay of the spectrum of the free running MLL (black solid curve) with the resonant wavelengths of rings 1 and 2 at tuning points P$_A$, P$_B$ and P$_C$. Labels indicate the resonant wavelengths of Ring 1. Ring 2 has the same bias point for P$_A$, P$_B$ and P$_C$ that corresponds to the central of the three wavelengths labeled as P$_B$.**

It is apparent that the next higher resonance, on the longer wavelength side, does not interact with MLL comb lines. On the other hand, the next lower resonant wavelength remains inside the MLL spectrum and interacts with weak comb lines. The data is summarized in the following table:

| Bias Point | $\lambda_1$ | $P_1$ | $\lambda_1{}'$ | $P_1{}'$ | $\lambda_2$ | $P_2$ | $\lambda_2{}'$ | $P_2{}'$ |
|---|---|---|---|---|---|---|---|---|
| $P_A$ | 1543.48 nm | -30.12 dBm | 1534.78 nm | -50.24 dBm | 1543.77 nm | -30.50 dBm | 1535.07 nm | -45.67 dBm |
| $P_B$ | 1543.77 nm | -30.50 dBm | 1535.07 nm | -45.67 dBm | 1543.77 nm | -30.50 dBm | 1535.07 nm | -45.67 dBm |
| $P_C$ | 1544.06 nm | -29.52 dBm | 1535.36 nm | -43.74 dBm | 1543.77 nm | -30.50 dBm | 1535.07 nm | -45.67 dBm |

wherein $\lambda_1$ and $\lambda_2$ refer to the main resonance wavelengths of rings 1 and 2, in the center of the MLL spectrum, $\lambda_1{}'$ and $\lambda_2{}'$ refer to the next lower resonances, and $P_1$, $P_2$, $P_1{}'$ and $P_2{}'$ refer to the power of the corresponding MLL lines. As a coincidence, the next lower resonances are also very closely aligned to an MLL line when the main resonance is aligned. It should also be noted that the MLL line power levels reported here should not be taken as absolute values, but just compared to each other, as the MLL spectrum was taken with high insertion losses (recorded at the output of the SiP chip with the laser at a large distance to the chip to minimize feedback).

It can be seen that the worst case is for Ring 1 at point $P_C$, at which the power of the MLL line at the next lower resonance was 14.2 dB below that of the main resonance. The best case is $P_A$, at which this number for ring 1 is 20.1 dB. From Fig. 8 of the main manuscript, this extra extinction does not appear to play a discernable role in the obtained results.

## IV. Derivation of Equations 1 and 2 from Main Manuscript

For the case of broadband reflection, we first derive the required conditions for the phase of all reflected lines to remain within a phase error of $\pm\pi/4$ relative to the reflection phase of a comb line at the center of the MLL's spectrum (that can be tuned with a phase tuner assumed to apply the same differential phase delay to all the comb lines). $\beta_0(\lambda)$ is the wave number of an ideal waveguide without any fabrication errors or dispersion, with a length resulting in a group delay that is a multiple of the laser repetition time. $\beta(\lambda)$ is the wave number of the actually fabricated waveguide, that has both finite dispersion and a fabrication induced group index change. Both depend on the wavelength $\lambda$. The MLL emission has a spectral width $\Delta\lambda$. $L$ is the round-trip length of the delay line, $c_0$ the speed of light in vacuum, $\omega$ the angular frequency of the light. $n_g$ and $n_{g0}$ are the group indices of the fabricated and of the nominal waveguide and $\Delta n_g = n_g - n_{g0}$ is the fabrication induced error. $v_g$ and $v_{g0}$ are the group indices and $D$ the dispersion parameter. This requirement is then translated into

$$L \cdot \left| \frac{d\beta - d\beta_0}{d\lambda} \left( \frac{\Delta\lambda}{2} \right) + \frac{1}{2} \frac{d^2\beta - d^2\beta_0}{d\lambda^2} \left( \frac{\Delta\lambda}{2} \right)^2 \right| < \pi/4$$

As we will see below, the first term in the equation above depends primarily on fabrication tolerances applied to the group index, and the second on the dispersion parameter of the waveguide. The error depends on $\beta - \beta_0$ rather than on $\beta$, since the ideal waveguide results in zero error. The group delay of the ideal waveguide is chosen for the phase error to be zero for all comb frequencies, so that only the deviation from this condition has to be considered.

Evaluating first the constraints on group index variation, we start with

$$L \cdot \left| \frac{d\beta - d\beta_0}{d\lambda} \left( \frac{\Delta\lambda}{2} \right) \right| < \pi/4$$

reshaped into

$$L \cdot \left| \frac{d\beta - d\beta_0}{d\omega} \frac{2\pi c_0}{\lambda^2} \left( \frac{\Delta\lambda}{2} \right) \right| < \pi/4$$

$$L \cdot \left| \frac{n_g - n_{g0}}{c_0} \frac{2\pi c_0}{\lambda^2} \left( \frac{\Delta\lambda}{2} \right) \right| < \pi/4$$

which then results into Eq. (1)

$$|\Delta n_g| < \frac{\lambda_0^2}{4L\Delta\lambda}$$

We now derive the second equation starting from

$$L \cdot \left| \frac{1}{2} \frac{d^2\beta - d^2\beta_0}{d\lambda^2} \left( \frac{\Delta\lambda}{2} \right)^2 \right| < \pi/4$$

which is reshaped into

$$L \cdot \left| \frac{1}{2} \frac{d}{d\lambda} \left( \left( \frac{1}{v_g} - \frac{1}{v_{g0}} \right) \frac{2\pi c_0}{\lambda^2} \right) \left( \frac{\Delta\lambda}{2} \right)^2 \right| < \pi/4$$

Since here we are investigating the main effect of waveguide dispersion instead of fabrication error, we assume $v_g(\lambda_0) = v_{g0}$ so that the condition can be further simplified into

$$L \cdot \left| \frac{1}{2} D \frac{2\pi c_0}{\lambda^2} \left( \frac{\Delta\lambda}{2} \right)^2 \right| < \pi/4$$

which finally results in Eq. (2)

$$|D| < \frac{\lambda_0^2}{Lc_0\Delta\lambda^2}$$

It now remains to be shown that Eqs. (1) and (2) also apply to the case of a ring based intra-cavity filter. In the following, $L$ is now the roundtrip length of the intra-cavity waveguides (including the effective delay length of the intra-cavity filter) and determines the FSR of the laser given by

$$FSR = \frac{\lambda_0^2}{Ln_g(\lambda_0)}$$

Note that only the waveguide group index at $\lambda_0$ determines the spectral offset between comb lines, as a consequence of mode-locking maintaining the comb-line to comb-line spacing constant. Moreover, if the length of the passive intra-cavity delay line is much longer than the length of the active waveguide, $n_g(\lambda_0)$ can be approximated as the group index of the passive delay line.

The maximum allowable linewidth of the intra-cavity filter scales with the FSR, since the filter needs to select comb lines from the neighboring ones. The linewidth in turn constrains the required spectral alignment of the filter's resonances. As a criterium, we set here that the detuning of the filter's passbands should be no more than an eighth of the laser's FSR. Given a group index offset $\Delta n_g$ (inside the filter) due to fabrication, the detuning $\delta\lambda$ at the edges of the MML spectrum is evaluated as

$$\delta\lambda = \frac{\Delta n_g}{n_g} \frac{\Delta\lambda}{2}$$

resulting in

$$\left| \frac{\Delta n_g}{n_g} \frac{\Delta\lambda}{2} \right| < \frac{\lambda_0^2}{8Ln_g}$$

which is again converted into Eq. (1) under the assumption that the delay line and the filter are made out of identical waveguides (for the two group indices to cancel out, otherwise they would need to be propagated through the derivation)

$$|\Delta n_g| < \frac{\lambda_0^2}{4L\Delta\lambda}$$

The detuning due to dispersion is evaluated as

$$\delta\lambda = \int_0^{\frac{\Delta\lambda}{2}} \frac{\Delta n_g(\lambda)}{n_g} d\lambda$$

Since $D = d(n_g/c_0)/d\lambda$

$$\delta\lambda = \int_0^{\frac{\Delta\lambda}{2}} \frac{c_0 D(\lambda - \lambda_0)}{n_g} d\lambda = \frac{c_0 D \Delta\lambda^2}{8n_g}$$

Thus we obtain

$$\left| \frac{c_0 D \Delta\lambda^2}{8n_g} \right| < \frac{\lambda_0^2}{8Ln_g(\lambda_0)}$$

resulting again in Eq. (2)

$$|D| < \frac{\lambda_0^2}{Lc_0\Delta\lambda^2}$$